\documentclass[reprint,amsmath,amssymb,aip,jcp]{revtex4-1}
\usepackage{flushend}
\usepackage{dcolumn}
\usepackage{bm}
\usepackage[version=4]{mhchem} 

\usepackage{xcolor, soul}
\usepackage{pdfcomment}
\usepackage{rotating}

\makeatletter
\def\@email#1#2{%
	\endgroup
	\patchcmd{\titleblock@produce}
	{\frontmatter@RRAPformat}
	{\frontmatter@RRAPformat{\produce@RRAP{*#1\href{mailto:#2}{#2}}}\frontmatter@RRAPformat}
	{}{}
}%

\makeatother
\usepackage{hyperref}
\begin{document}
	
\preprint{AIP/123-QED}


\title{Group-I lead oxide  X$_2$PbO$_3$ (X=Li, Na, K, Rb, and Cs) glass-like materials for energy applications: A hybrid-DFT study}

\author{R. Zosiamliana}
\affiliation{Department of Physics, Mizoram University, Aizawl-796009, India}
\affiliation{Physical Sciences Research Center (PSRC), Department of Physics, Pachhunga University College, Mizoram University, Aizawl-796001, India}

\author{Lalhriat Zuala}
\affiliation{Physical Sciences Research Center (PSRC), Department of Physics, Pachhunga University College, Mizoram University, Aizawl-796001, India}

\author{Lalrinthara Pachuau}
\affiliation{Physical Sciences Research Center (PSRC), Department of Physics, Pachhunga University College, Mizoram University, Aizawl-796001, India}

\author{Lalmuanpuia Vanchhawng}
\affiliation{Physical Sciences Research Center (PSRC), Department of Physics, Pachhunga University College, Mizoram University, Aizawl-796001, India}

\author{S. Gurung}
\affiliation{Physical Sciences Research Center (PSRC), Department of Physics, Pachhunga University College, Mizoram University, Aizawl-796001, India}

\author{A. Laref}
\affiliation{Department of Physics and Astronomy, College of Science, King Saud University, Riyadh, 11451, Saudi Arabia}%

\author{D. P. Rai}
\email{dibyaprakashrai@gmail.com}
\affiliation{Department of Physics, Mizoram University, Aizawl-796004, India}%
\date{\today}
	
\begin{abstract}
Pb-based compounds have garnered considerable theoretical and experimental attention due to their promising potential in energy-related applications. In this study, we explore the glass-like alkali metal lead oxides \texorpdfstring{X$_2$PbO$_3$}{X2PbO3} (X=Li, Na, K, Rb, Cs) and assess their suitability for piezoelectric and thermoelectric applications. First-principles calculations were performed using hybrid density functional theory (DFT), incorporating B3LYP, HSE06, and PBE0 functionals. Among these, PBE0 is identified as the most accurate, yielding lattice parameters in close agreement with experimental data. Structural stability was confirmed through evaluation of thermal, mechanical, and formation energies. For the non-centrosymmetric orthorhombic phase C$mc$2$_1$-\texorpdfstring{X$_2$PbO$_3$}{X2PbO3} (X=K, Rb, Cs), piezoelectric constants were computed via both the numerical Berry phase (BP) method and the analytical Coupled Perturbed Hartree-Fock/Kohn-Sham (CPHF/KS) formalism. Notably, \texorpdfstring{Cs$_2$PbO$_3$}{Cs2PbO3} exhibited a piezoelectric coefficient of \texorpdfstring{e$_{33}$}{e33} = 0.60 C m$^{-2}$ (CPHF/KS), while \texorpdfstring{K$_2$PbO$_3$}{K2PbO3} showed \texorpdfstring{e$_{32}$}{e32} = -0.51 C m$^{-2}$ (BP). Thermoelectric properties were investigated using the semiclassical Boltzmann transport theory within the rigid band approximation. The calculated thermoelectric performance reveals promising figures of merit (ZT), ranging from 0.3 to 0.63, suggesting these materials are applicable as future thermoelectric materials.
\end{abstract}

\maketitle
	\section{Introduction}
	\label{Introduction}
Energy generation and storage is crucial to meet the current global crisis. The energy extracted from the fossil fuels are extensively utilized for transportation, many industries and may other purposes, despite its known harmful effect.\cite{Martins2021b} Thermoelectricity holds promise as suitable and sustainable alternative to greenhouse gases emitting fuels if efficient thermoelectric (TE) device of high heat energy-to-electric voltage conversion efficiency can be discovered.\cite{Tritt2006b,Freer2020,Soleimani2020,Finn2021,Singh2024} First-principles calculations with the incorporation of Boltzmann transport equation via BoltzTraP code\cite{Madsen2006a} can help in predicting the the candidates for the preparation of TE devices. Glass-like compounds have become a fascinating materials; both experimentally and theoretically due to their complexity of structures which inturn exhibit low lattice thermal conductivity, tunable electronic properties, and flexibility in fabrication.\cite{Holomb2013b,Gong2021b,Biskri2014a,Du2006c} Although silicate (SiO$_2$) glasses are widely abundant and commonly used materials have significant limitations such as brittleness, high resistance and phase transition at higher temperatures, etc.  Therefore, it is pivotal for researchers to find proper replacement of silicate glasses.\cite{Renthlei2023d} In this novel, we propose alkali metal oxide Pb-based glass-like materials X$_2$PbO$_3$ (X=Li, Na, K, Rb, Cs) as a potential replacement for SiO$_2$-glasses such as: Li$_2$SiO$_3$, Na$_2$SiO$_3$, etc., and conduct a thorough investigation on these compounds using the density functional theory (DFT), particularly for their TE applications. To the best of our knowledge, among the suggested alkali metal oxides: the K$_2$PbO$_3$ was the first compound to be synthesized back in 1964 by Hoppe and co-workers,\cite{Hoppe1964} later in the year 1972 the Cs$_2$PbO$_3$ compound was synthesized by Panek \textit{et al.},\cite{Panek1972} and then latterly the Rb$_2$PbO$_3$ was again synthesized in 1977 by Hoppe and St{\"o}ver.\cite{Hoppe1977} From these earlier experimental studies, it was found that the K$_2$PbO$_3$, Rb$_2$PbO$_3$, and Cs$_2$PbO$_3$ existed in orthorhombic structure with space group C$mc$2$_1$ (No. 36). More recently, in 1982, Brazel and Hoppe synthesized the Li$_2$PbO$_3$ by decomposing K$_2$Li$_6$[Pb$_2$O$_8$] and reported the monoclinic structural symmetry (C2/c space group) for this compound.\cite{Brazel1982a} The entertaining information about the C$mc$2$_1$ symmetry materials of X$_2$PbO$_3$ is that in these structures, similar to the prototype Na$_2$SiO$_3$, and Na$_2$GeO$_3$, the presence of link between Oxygen and Pb-atoms led to the formation of the three dimensional network of tetrahedral chain of [PbO$_4$]. The link forms the bridge-oxygen (BO) and non-bridge-oxygen (NBO), where alkali atoms such as X=K, Rb, and Cs are bonded.\cite{Zachariasen1932b} Therefore, K$_2$PbO$_3$ (KPO), Rb$_2$PbO$_3$ (RPO), and Cs$_2$PbO$_3$ (CPO) are non-centrosymmetric in structure, and consequently exhibit piezoelectric properties. Moreover, in case of Li$_2$PbO$_3$ (LPO) and Na$_2$PbO$_3$ (NPO), since the compounds are centrosymmetric, they do not possess piezoelectricity.
\par From past few decades, the experimental and theoretical insight into novel the multi-function energy materials especially for TE and piezoelectric applications have become a hot topic among researchers.\cite{Mezilet2022,Sadouki2023,Zhang2022} The thermoelectricity and piezoelectricity hold the requisite hallmarks for green energy resources as they produce electricity without any emissions of toxic pollutants. The glass-like materials being possessing high mechanical and thermal stability, low lattice thermal conductivity ($\kappa_L$), high melting temperature (T$_m$), ease of availability, and the cost-effectiveness have made these materials a perfect candidate for multi-use energy generators. However, the presence of toxic Pb-element in our proposed compounds i.e., X$_2$PbO$_3$ have made this topic more challenging. From various research article, it has been observed that the toxicity offered by Pb-element can be successfully minimized, and the prominent strategy among them are chelation therapy, nano-encapsulation, and N-acetylcysteine (NAC).\cite{McKay2013,Flora2013,Yedjou2007} To reduce the energy crisis rendered by the global demands of energy sources, thermoelectricity and piezoelectricity could be an innovative approach although their efficiencies are low. From the recent study led by Zosiamliana and co-workers,\cite{Renthlei2023e} the reported TE efficiencies at T=1200 K for Pb-based perovskites such as PbTiO$_3$, PbZrO$_3$, and PbHfO$_3$ were ZT=0.64, 0.66, and 0.61, revealing the suitability of Pb-based materials to TE devices. Also, recent report reveals the relevancy of glass-like materials such as Na$_2$SiO$_3$ and Na$_2$GeO$_3$ (the prototype compounds for KPO, RPO, and CPO) for piezoelectric devices, with a response of e$_{33}$=0.22 C m$^{-2}$ and e$_{33}$=0.91 C m$^{-2}$, respectively.\cite{Renthlei2023d,Zosiamliana2022g,Zosiamliana2022h} Experimentally and theoretically, only few research have been carried out on glass-like materials for TE applications. Recently, an efficiency of ZT=0.027 at T=393 K was reported for Na$_2$SiO$_3$ as Graphite/mixture (CuSO$_4$+CoOH+SiO$_2$+Na$_2$SiO$_3$) /Aluminum by Chira \textit{et al}.\cite{Chira2020a} Thus, the unmet research area needs to be addressed is the piezoelectric and TE applications of X$_2$PbO$_3$ glass-like materials.

\par As far as we know, from several surveyed literature, only few experimental and theoretical studies were conducted for these materials. Recently, the photo-catalytic use of these compounds for solar-to-hydrogen conversion was reported by Gelin and colleagues\cite{Gelin2024} however, failed to provide a comprehensive insight into the fundamental properties. As a result, the main focus of this work will be on the thorough analysis of the fundamental characteristics of X$_2$PbO$_3$ (X=Li, Na, K, Rb, Cs) and their applications for piezoelectric and TE materials employing hybrid-DFT.

\section{Computational Details}
In this work we have performed the density functional theory (DFT) as implemented in the CRYSTAL17-code to evaluate the physical properties of X$_2$PbO$_3$. In this part of DFT frame work the crystal orbitals are described by the linear combination of Gaussian-type functions (GTF).\cite{Dovesi2018b} The atomic centers for all the constituent atoms such as X=Lithium (Li), Sodium (Na), Potassium (K), Rubidium (Rb), and Caesium (Cs), Lead (Pb), and Oxygen (O) were described by a revised triple-$\zeta$ valence plus polarization (TZVP) basis set.\cite{VilelaOliveira2019b,Laun2022a} In this work, four different exchange-correlation functionals were adopted, namely; (1) Perdew-Burke-Ernzerhof (PBE) generalized gradient approximation (GGA);\cite{Perdew1996h} Global hybrids: (2) Becke 3-parameter Lee-Yang-Parr (B3LYP),\cite{Becke1988a,Lee1988a} and (3) PBE with 25\% Fock exchange (PBE0);\cite{Adamo1999a} and (4) range-separated hybrid screened-Coulomb (SC) called HSE06.\cite{Heyd2003a} 
\par The expression for the employed hybrid functionals are:
\begin{equation}
\begin{split}
E^{B3LYP}_{XC}=E^{LSDA}_{XC}+a_0(E^{exact}_{X}-E^{LSDA}_{X})+a_X\Delta E^{B88}_{X}
\\+a_C\Delta E^{PW91}_{C}
\end{split}
\end{equation} 
\par Here, a$_0$, a$_X$, and a$_C$ are semi-empirical coefficients, E$^{LSDA}_{XC}$ is the local spin density exchange correlation, E$^{exact}_{X}$ is the exact exchange energy, $\Delta$E$^{B88}_{X}$ is Becke's gradient correction for exchange, and $\Delta$E$^{PW91}_{C}$ is the Perdew and Wang gradient correction.
\begin{equation}
E^{PBE0}_{XC}=\frac{1}{4}E^{HF}_{X}+\frac{3}{4}E^{PBE}_{X}+E^{PBE}_{C}
\end{equation} 
\par Here, the full PBE correlation energy and the 3:1 ratio of PBE to HF exchange energies determine the PBE0 functional.
\begin{equation}
\begin{split}
E^{HSE06}_{XC}=\frac{1}{4}E^{SR, HF}_{X} (\omega)+\frac{3}{4}E^{SR, PBE}_{X} (\omega)+E^{LR, PBE}_{X} (\omega)
\\+E^{PBE}_{C} 
\end{split}
\end{equation} 
\par Where, $\omega$=0.2 is the screening parameter. And E$^{SR, HF}_{X}$ represents the short-range Hartree-Fock exact exchange functional, E$^{SR, PBE}_{X}$, and E$^{LR, PBE}_{X}$ are the short-range, and long-range PBE exchange functionals, respectively, and E$^{PBE}_{C}$ is the full correlation energy. 
\par For the structural optimizations, an analytical quasi-Newtonian appraoch combined with Hessian Broyden-Fletcher-Goldfarb-Shanno (BFGS) scheme was used.\cite{Head1985,Nawi2006} To check the convergence, the gradient components and nuclear displacements with tolerances on their root-mean-square (RMS) were set to 0.0001 and 0.0004 Hartree (Ha), respectively. The accuracy of convergence criteria was set to the five thresholds which control the overlap and penetration of Coulomb integrals, the overlap for HF exchange integrals, and the pseudo-overlap were set to (10$^{-7}$, 10$^{-7}$, 10$^{-7}$, 10$^{-7}$, 10$^{-7}$, 10$^{-14}$). The first Brillouin zone integration was performed using 10$\times$10$\times$10 k-mesh within the Monkhorst-Pack scheme,\cite{Monkhorst1976g} and an energy convergence criteria of 10$^{-7}$ Ha was considered. For the calculation of the physcial properties we have set the higher k-mesh of 12$\times$12$\times$12. For the calculation of elastic and piezoelectric properties, we have opted the best suited exchange correlation functionals depending on the accuracy in reproducing the experimental lattice parameters. To verify the structural stability, we performed the ab initio molecular dynamics (AIMD) simulation with nVT canonical ensemble (calculation details are provided in Section \ref{SPS}).\cite{NosE2002,Andersen1980a,Smidstrup2019a} 
\par For TE properties calculation, Boltzmann transport semi-classical equation (BTE) within a rigid band approximation (RBA) using BoltzTraP as implemented in CRYSTAL17-code was employed.\cite{Madsen2006a,Scheidemantel2003a,Ryu2016a} The wave functions were recalculated at a dense k-mesh of 42$\times$42$\times$42 within the first Brillouin zone. A constant relaxation time approximation (CRTA) for carriers was assumed for all the investigated compounds, and fixed it at $\tau$=10$^{-14}$ s (default $\tau$ for BoltzTraP-code).
	
	\section{Results and Discussions}
	
	\subsection{Structural Properties and Stability}
	\label{SPS}
	\begin{figure*}[hbt!]
		\includegraphics[height=6.2cm]{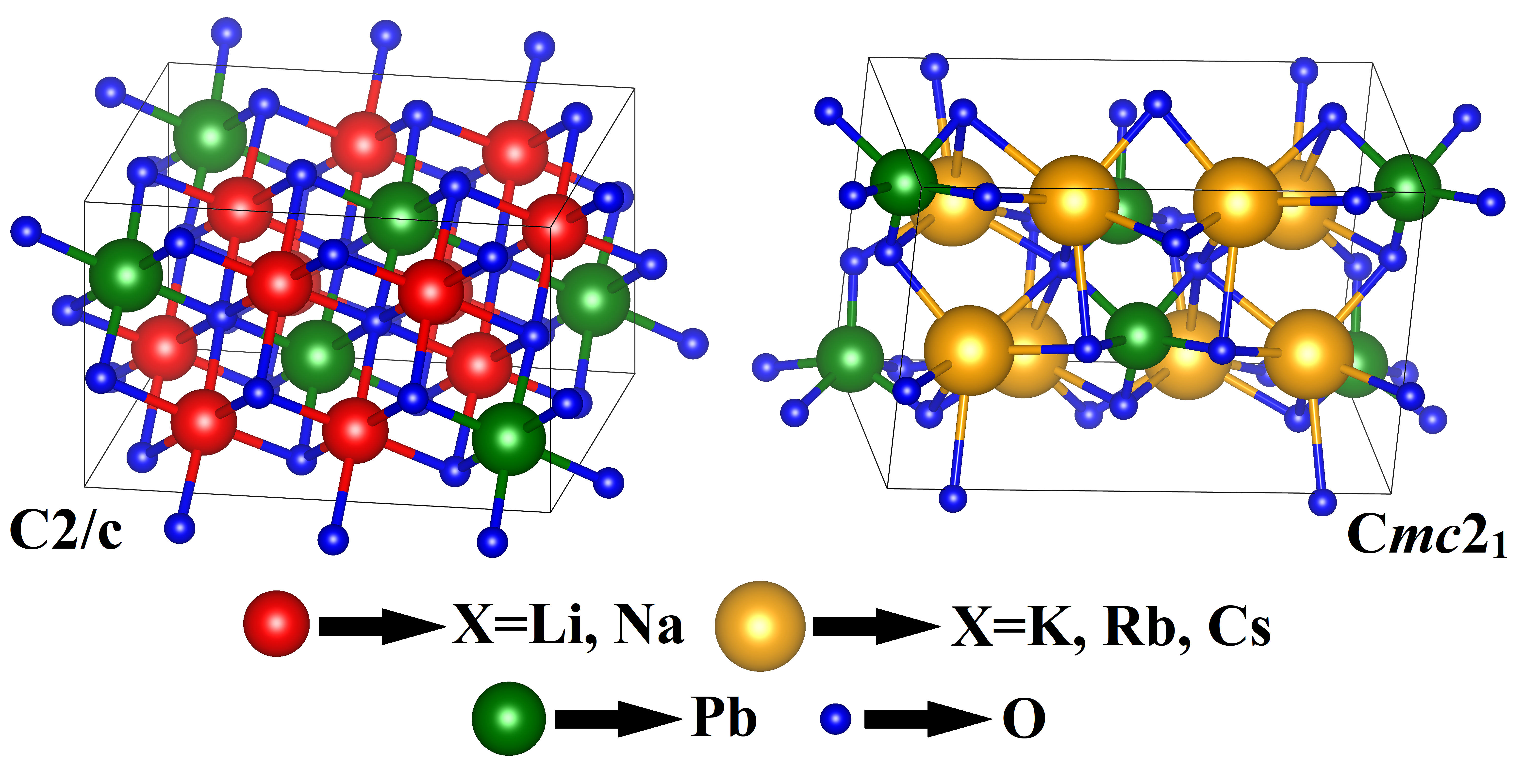}
		\caption{The C2/c (X=Li, Na), and C$mc$2$_1$ (X=K, Rb, Cs) structures of X$_2$PbO$_3$ viewed using an external program called VESTA.\cite{Momma2008a}}
		\label{structure}
	\end{figure*}
	The structural optimization results using various adopted functionals such as PBE-GGA, B3LYP, HSE06, and PBE0 reveal that the investigated compounds X$_2$PbO$_3$ crystallized in monoclinic symmetry (C2/c space group) for X=Li, and Na, and orthorhombic structure (C$mc$2$_1$ space group) for X=K, Rb, and Cs [see figure \ref{structure}]. The agreements between the calculated lattice constants with the available experimental data for each compounds are presented in table S1. For LPO, KPO, RPO, and CPO the global hybrid-PBE0 functional is found to be the most relevant functional in reproducing the experimental data with $\lvert$$\Delta$a$\rvert$, $\lvert$$\Delta$b$\rvert$, and $\lvert$$\Delta$c$\rvert$ $<$ 2\%, as it is known that global hybrid-PBE0 functional reproduces better lattice parameters and the electronic band structure of small or large band gap solids. It is important to remain aware that the stability of the studied systems will be impacted if the lattice constant's precision deviates by more than 2\%.\cite{Renthlei2023e} From the formation energy (E$^f$) calculations using equation \ref{For}, negative E$^f$ revealed the ground state structural stabilities for all the considered systems in their corresponding phases, and implies to the realization of their experimental synthesis. Furthermore, greater negative E$^f$ is observed for each compound with PBE0 functional, suggesting this functional is the most favored to acquire the energy ground state. As a result, each compound's optical, elastic, and piezoelectric properties are investigated using the PBE0 functional. However, the electronics and TE properties are explicitly calculated using the four different adopted functionals in order to verify the role played by exchange correlation functions during the study of TE properties.    
	\begin{equation}
	\label{For}
	E^f=\frac{1}{12}\big[E_{tot}-\big(4E_{X}+2E_{Pb}+6E_{O}\big)\big]
	\end{equation}
	\par
\begin{figure}[t!]
		\includegraphics[height=9.0cm]{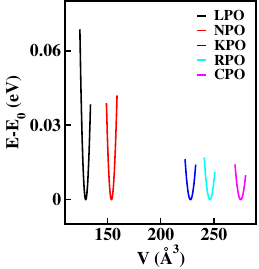}
		\caption{Energy vs volume curves of X$_2$PbO$_3$ (X=Li, Na, K, Rb, Cs) calculated using Birch-Murnaghan's EOS curve fitting method. Here, E$_0$ is ground state energy.}
		\label{lattice}
	\end{figure}

 Here, E$_{tot}$ is total ground state energy, and E$_X$, E$_{Pb}$, and E$_O$ are corresponding single atom ground state energy for X, Pb, and O atoms. Since, there are 12 atoms in the unit cell, the right hand side (RHS) of equation \ref{For} is divided by 12. To obtain the most stable configurations for all the examined systems, total energy versus the unit cell volumes were fitted through the Birch-Murnaghan equation of states (EOS) scheme given by equation \ref{EOS} using PBE0 functional.\cite{Birch1947b,Murnaghan1937b} From the depicted smooth parabolic curves in figure \ref{lattice}, the energy difference where E-E$_0$=0 eV corresponds to the most stable structure. 
	\begin{equation}
	\label{EOS}
	\begin{split}
	E(V)=E{_0}+\frac{9{\times}B{_0}V{_0}}{16} [{ [(\frac{V_0}{V})^\frac{2}{3}-1]{^3}{\times}B'{_0}}+[(\frac{V_0}{v})^\frac{2}{3}\\
	-1]^2{\times}[6-4\times(\frac{V_0}{V})^\frac{2}{3}]]
	\end{split}
	\end{equation}    
	\begin{figure*}[t!]
		\includegraphics[height=12.0cm]{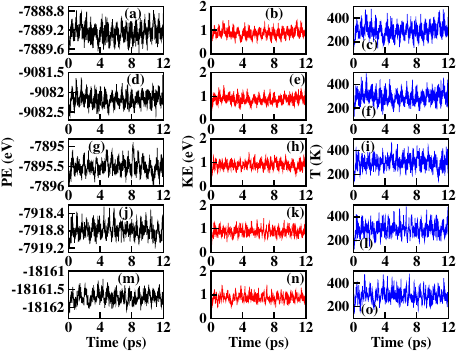}
		\caption{The potential energy (PE), kinetic energy (KE), and evolution of temperature (T) as a function of time steps calculated using MD-simulation based on nVT-canonical ensemble: (a-c) X=Li, (d-f) X=Na, (g-i) X=K, (j-l) X=Rb, and (m-o) X=Cs.}
		\label{md}
	\end{figure*} 
	\par To verify the thermal stabilities of the studied systems, we have performed the molecular dynamics (MD) simulation for each relaxed conventional cell structures (i.e., two times the number of atoms in the unit cell). Since, MD-simulation cannot be performed using the CRYSTAL17-code, therefore, throughout this computational process, the QuantumATK Q-2019.12 code which rely on a linear combination of the atomic orbital method (LCAO) with the canonical ensemble (nVT) based on Nos\'e-Hoover thermostat was adopted.\cite{Parrinello1981a,Parrinello1980b,Nose1990b} A 12 ps total simulation time with 4 fs time step was considered during the simulations for all the structures. The graphical representation of MD-simulation [see figure \ref{md}] calculated at room temperature (T=300 K) and the simulations in nVT ensemble where the number of atoms, cell volume, and temperature are constant provide the prediction of more realistic insight into the evolution of energies (PE, and KE), and temperature (T). Here, PE and KE represent the sum of energies from non-bonded and bonded interactions, and the heat absorbed by the systems, respectively. The higher evolution of T with time suggested more movement of the atoms, thereby resulting in more fluctuation of KE. The nearly linear fluctuations of PE, KE, and T even upto 12 ps time steps reveals the thermal stability for all the investigated materials. 
	
	\subsection{Electronic Properties}
	\label{Electronic Properties}
\begin{figure*}[t!]
		\includegraphics[height=12.0cm]{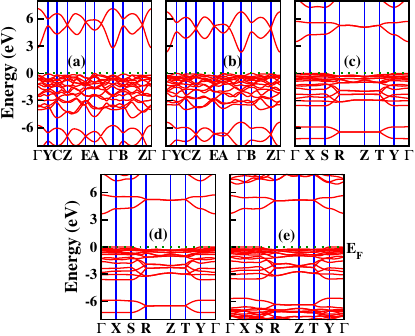}
		\caption{Calculated band structures for X$_2$PbO$_3$ using PBE0 functional: (a) X=Li, (b) X=Na, (c) X=K, (d) X=Rb, and (e) X=Cs.}
		\label{band}
	\end{figure*}

\begin{figure*}[t!]
		\includegraphics[height=12.0cm]{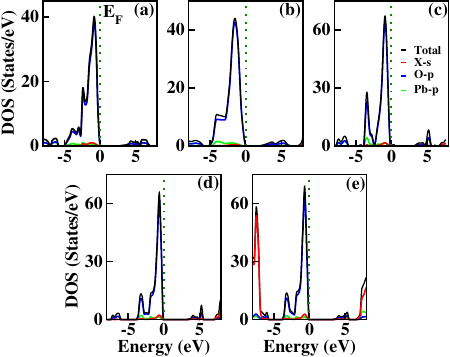}
		\caption{Calculated DOS for X$_2$PbO$_3$ using PBE0 functional: (a) X=Li, (b) X=Na, (c) X=K, (d) X=Rb, and (e) X=Cs.}
		\label{dos}
	\end{figure*}

	The emergence of physical properties can be understood by interpreting electronic properties with regard to orbital interactions at the atomic level. In this section, we have studied the electronic properties which includes band structures, density of states (DOS), and atomic charge transfer (Q$_T$) from PBE-GGA, B3LYP, HSE06, and PBE0 functionals. From the presented band structures, and DOS calculated using PBE0 functional in figures \ref{band}, and \ref{dos} (for PBE-GGA, B3LYP, and HSE06 see figures S1, and S2), and the energy gap (E$_g$) at the high symmetry points given in table \ref{EPT-1} revealed the indirect semi-conducting band gap nature for LPO, and NPO with top and bottom of valence and conduction bands lie at A, and $\Gamma$-symmetry points. While for KPO, RPO, and CPO, E$_g$ are along $\Gamma$-symmetry suggesting a direct band gap semiconductor behavior. The incorporation of hybrid functional flavors during the electronic properties calculations have significant effect in the E$_g$ due to the shifts in the energy levels, however, the energy band profiles are preserved. Comparing the obtained E$_g$ with the recent report led by Gelin \textit{et al.},\cite{Gelin2024} where E$_g$ were recorded for C2/c symmetry of Li$_2$PbO$_3$, P6$_3$/$mmc$ symmetry of K$_2$PbO$_3$, P$nma$ symmetry of Rb$_2$PbO$_3$, and C$mc$2$_1$ symmetry of Cs$_2$PbO$_3$ using GGA, and GGA+U approximation, one can find that these reported E$_g$ accord well with our results. 
	
\begin{table*}[hbt!]
		\small
		\caption{Electonic band gap (E$_g$) (in eV), and atomic charge transfer (Q$_T$) (in $\lvert$e$\rvert$) using Mulliken population analysis for X$_2$PbO$_3$ (X=Li, Na, K, Rb, Cs).}
		\label{EPT-1}\renewcommand{\arraystretch}{1.75}
		\begin{tabular*}{1.5\textwidth}{|l|lllll|lllll|lllll|lllll|}
			\hline
			\multicolumn{1}{l}{} & &&\multicolumn{3}{l}{PBE-GGA} &&&\multicolumn{3}{l}{B3LYP} &&&\multicolumn{3}{l}{HSE06}&&&\multicolumn{3}{l}{PBE0}\\
			\hline
X & {E$_g$} & {Q$^T_X$} & {Q$^T_{Pb}$} & {Q$^T_{BO}$} & {Q$^T_{NBO}$} & {E$_g$} & {Q$^T_X$} & {Q$^T_{Pb}$} & {Q$^T_{BO}$} & {Q$^T_{NBO}$} & {E$_g$} & {Q$^T_X$} & {Q$^T_{Pb}$} & {Q$^T_{BO}$} & {Q$^T_{NBO}$} & {E$_g$} & {Q$^T_X$} & {Q$^T_{Pb}$} & {Q$^T_{BO}$} & {Q$^T_{NBO}$} \\
			\hline
			Li  & 0.88 & 0.54 & 1.71 & -0.95 & -0.92 & 2.27 & 0.65 & 1.99 & -1.10 & -1.09 & 2.23 & 0.61 & 2.01 & -1.08 & -1.07 & 2.82 & 0.61 & 2.04 & -1.09 & -1.08 \\
			Na  & 0.53 & 0.74 & 1.49 & -1.01 & -0.98 & 1.83 & 0.82 & 1.76 & -1.15 & -1.12 & 1.79 & 0.81 & 1.77 & -1.15 & -1.12 & 2.34 & 0.82 & 1.80 & -1.16 & -1.13 \\
			K   & 1.23 & 0.71 & 1.34 & -0.93 & -0.92 & 2.94 & 0.75 & 1.55 & -1.01 & -1.01 & 2.84 & 0.75 & 1.57 & -1.03 & -1.03 & 3.50 & 0.75 & 1.59 & -1.03 & -1.03 \\
			Rb  & 1.45 & 0.77 & 1.27 & -0.95 & -0.93 & 3.08 & 0.81 & 1.47 & -1.03 & -1.02 & 3.02 & 0.82 & 1.50 & -1.05 & -1.04 & 3.67 & 0.82 & 1.51 & -1.05 & -1.04 \\
			Cs  & 1.53 & 0.77 & 1.24 & -0.95 & -0.92 & 3.13 & 0.82 & 1.43 & -1.03 & -1.02 & 3.09 & 0.81 & 1.47 & -1.03 & -1.02 & 3.72 & 0.81 & 1.48 & -1.03 & -1.03 \\
			\hline
		\end{tabular*}
	\end{table*}

	\par Insight into the general distribution of electronic states at different energy levels and the electronic charge distributions upon formation of bonds	are gained through the analysis of DOS and Mulliken population analysis.\cite{Mulliken1955} From the DOS plots, it is comprehensible that for all the studied compounds, the main contributing states near the Fermi level (E$_F$) along the valence band region are from the O-p states, while the major contribution along the conduction region are from the complex hybridized states of X-s, Pb-p, and O-p, which are analogous to those closely related glass-like materials such as Li$_2$SiO$_3$, Na$_2$SiO$_3$, Li$_2$GeO$_3$, and Na$_2$GeO$_3$.\cite{Renthlei2023d,Zosiamliana2022g,Han2020a,Dien2021c} For each compound, from the relatively greater atomic concentrations of NBO than BO, and from the simple altercation of on-site Coulomb repulsion, the NBO-p orbitals which are with greater energy (or lower binding energy) are likely to have larger valence charges than BO-p orbitals. Therefore, among the two prominent DOS peaks along the valence region found between -5 eV and 0 eV, the first peak which is closer to the E$_F$ is mainly from NBO-p contributions, while the second peak which is at the lower energy region is contributed by BO-p. Furthermore, from the data of Mulliken population analysis of the relative atoms charge distribution calculated from different adopted functionals presented	in table \ref{EPT-1}, the X and Pb atoms transfer charges while	the NBO and BO atoms accumulate them for all the considered systems. By establishing a correlation between charge transfer and DOS, it is possible to clarify the impact of various functionals on the electronic properties calculation. For LPO, and NPO, from the DOS plots it is possible to notify that total DOS (TDOS) contribution near E$_F$ is comparatively higher with PBE-GGA functional when compared to other adopted functionals. This is mainly due to a significant change in Q$^T$ between PBE-GGA and hybrid functionals calculated results. With hybrid functionals the charge accumulated by BO and NBO becomes larger suggesting lower number of unoccupied states, which subsequently reduces the TDOS contribution around E$_F$. However, for KPO, RPO, and CPO an opposite phenomenon of PBE-GGA estimated TDOS exhibiting lower contribution near E$_F$ than those hybrid functionals employed are observed, even-though Q$^T$ for BO and NBO becomes larger with hybrid functionals. The structural arrangements of the materials can provide insight into the reason behind this such behavior. The X-BO and X-NBO bond lengths in C2/c symmetry compounds (i.e., X=Li and Na) are identical, while they are unequal in C$mc$2$_1$ symmetry compounds (i.e., X=K, Rb, and Cs). The difference in this structural arrangements of the materials for C2/c, and C$mc$2$_1$ symmetry compounds is the key factor for the non-centrosymmetric nature of KPO, RPO, and CPO compounds.      
	\par Figure S15 illustrates the two-dimensional differential charge density map for the investigated compounds, providing profound insights into the charge distribution among X-BO/NBO, Pb-BO/NBO, and X-Pb interactions. The visualization highlights charge depletion in red and charge accumulation in blue, reflecting the redistribution of electron density within the system. This charge rearrangement plays a crucial role in elucidating the relationship between charge transfer mechanisms and electronic band structures, particularly concerning electron transitions from the valence band to the conduction band through the electronic band gap. In each case, X and Pb atoms predominantly act as electron donors, leading to localized charge depletion in their vicinity. In contrast, oxygen atoms, due to their higher electronegativity, attract electron density, resulting in significant charge accumulation around O-sites. This pronounced electron localization at oxygen centers underscores their function as electron-rich species, thereby influencing the overall electronic structure and bonding nature of the studied materials. The observed charge density differences align well with other electronic properties, including the calculated density of states (DOS) and Mulliken population analysis. Furthermore, while no substantial variations are observed between the employed functionals, noticeable differences arise when comparing the C2/c-X$_2$PbO$_3$ and C$mc$2$_1$-X$_2$PbO$_3$ phases. Specifically, the C2/c-X$_2$PbO$_3$ structure exhibits a more pronounced presence of red and blue regions than the C$mc$2$_1$-X$_2$PbO$_3$ phase, indicating stronger charge transfer interactions. This suggests a more significant charge redistribution in the C2/c system, highlighting its enhanced electronic polarization and bonding characteristics.
	\subsection{Optical Properties}
	\begin{figure*}[hbt!]
		\includegraphics[height=7.00cm]{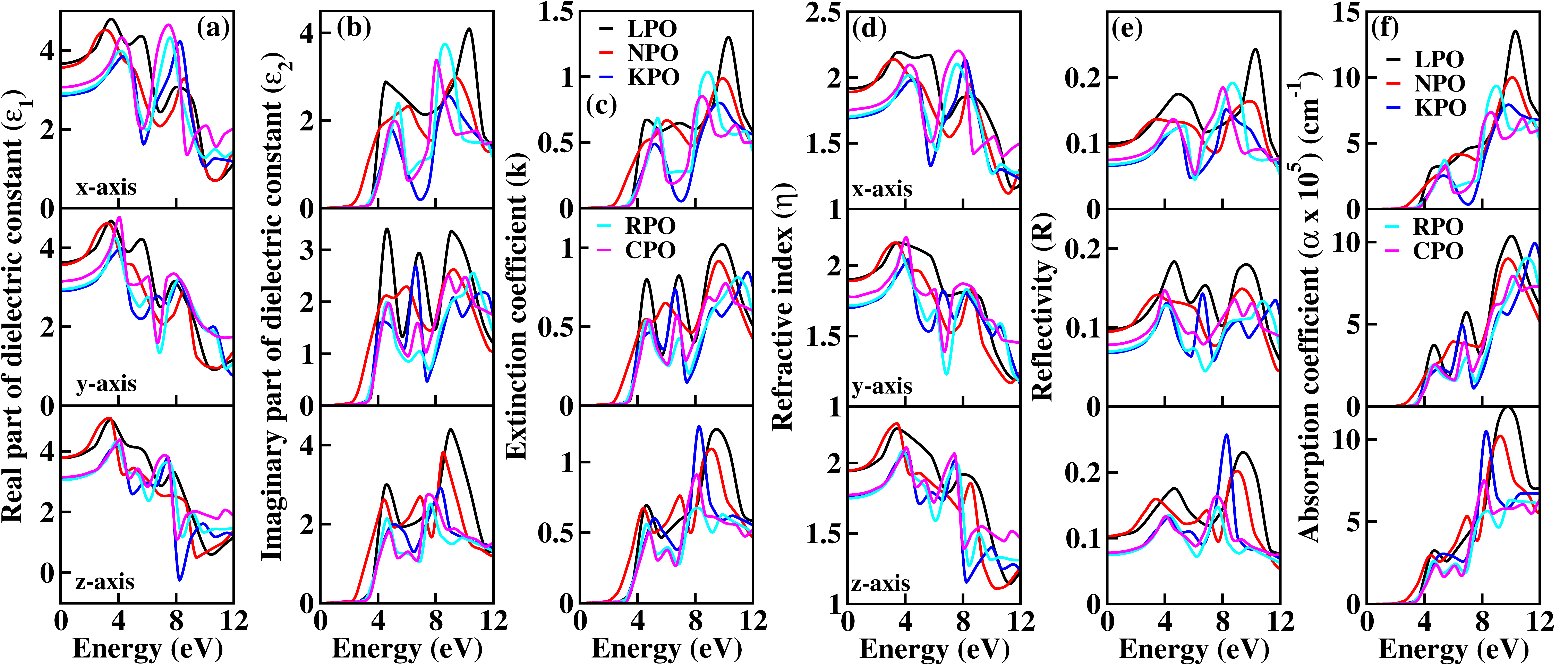}
		\caption{Calculated optical properties of X$_2$PbO$_3$ (X=Li, Na, K, Rb, Cs) using PBE0 functional: (a) real part of dielectric constant ($\epsilon_1$), (b) imaginary part of dielectric constant ($\epsilon_2$), (c) extinction coefficient (k), (d) refractive index ($\eta$), (e) reflectivity (R), and (f) absorption coefficient ($\alpha$).}
		\label{optical}
	\end{figure*}
	To interpret the interaction between electromagnetic radiation and the studied X$_2$PbO$_3$ materials, we have explored the optical properties by calculating the complex dielectric constants ($\epsilon$), extinction coefficient, refractive index, reflectivity, and absorption coefficient as a function of photon energy using PBE0 functional. It is known that at higher frequencies, $\epsilon$ is divided in two parts, namely; the real part ($\epsilon_1$), and the imaginary part ($\epsilon_2$) [see equation \ref{EqOP-1}].\cite{Ambrosch-Draxl2006d}
	\begin{equation}
	\epsilon=\epsilon_1+i\epsilon_2 
	\label{EqOP-1}
	\end{equation} 
	\par For calculating other optical parameters like extinction coefficient (k), refractive index ($\eta$), reflectivity (R), and absorption coefficient ($\alpha$), the employed formulae are:
	\begin{equation}
	k(\omega) = \sqrt{\frac{(\epsilon_1^2 + \epsilon_2^2)^\frac{1}{2} - \epsilon_1}{2}}
	\label{EqOP-2}
	\end{equation}
	\begin{equation}
	\eta(\omega) = \sqrt{\frac{(\epsilon_1^2 + \epsilon_2^2)^\frac{1}{2} + \epsilon_1}{2}}
	\label{EqOP-3}
	\end{equation}
	\begin{equation}
	R(\omega) = \frac{(1-\eta)^2+k^2}{(1+\eta)^2+k^2}
	\label{EqOP-4}
	\end{equation}
	\begin{equation}
	\alpha(\omega) = \frac{2\omega\kappa(\omega)}{c}
	\label{EqOP-5}
	\end{equation}
	\par The $\epsilon_1$ which is closely interrelated with $\eta$ determines the amount of material polarization and dispersion of electromagnetic radiation when interacting with the material surface [see figure \ref{optical} (a) and (d)]. The calculated static real dielectric constant $\epsilon_1$(0) are in the order of LPO $>$ NPO $>$ CPO $>$ RPO $>$ KPO, with values in the range of 2.80 to 3.90 arb. unit, for all x, y, and z-axes. Along the x-axis, the first prominent peaks for $\epsilon^x_1$ are found within 3.0 to 4.5 eV and undergo blue shifting as X goes from Na $\rightarrow$ Li $\rightarrow$ Cs $\rightarrow$ Rb $\rightarrow$ K. This indicates that the maximum probable interaction with electromagnetic radiation not only differ but also varies across frequency ranges for each compounds. Among all the investigated systems, higher static refractive indices $\eta$(0) values are observed for X=Li, and Na than when X=K, Rb, and Cs, with $\eta$(0) values ranging from 1.5 to 2.0, revealing that they are translucent in nature. Corresponding to $\epsilon_1$, the curves of $\eta$ show similar trends. The decreasing refractive indices at higher photon energy region i.e., beyond 8 eV reveals that the electromagnetic radiation no longer have sufficient energy to interact electronically with the material's electron. 
	\par The $\epsilon_2$ which is in close relation with the k and $\alpha$ corresponds to the inter-band transition between the valence and conduction bands. From figure \ref{optical} (b), on analyzing the directional wise interpretation of $\epsilon_2$, it is noticeable that $\epsilon_2$ is highly anisotropic in nature. Clearly, one can find two prominent peaks of $\epsilon_2$ for each compound along x and z-axes, and three distinct peaks in the y-axis. Along the x-axis, the first prominent peaks of $\epsilon^x_2$ are blue shifted as X moves from Li $\rightarrow$ K $\rightarrow$ Cs $\rightarrow$ Rb $\rightarrow$ Na. The extinction coefficient (k) which determines the materials capacity to absorb radiation of a specific wavelength is depicted in figure \ref{optical} (c). The obtained minimum threshold energy for X=Li, Na, K, Rb, and Cs are 3.51, 2.33, 3.46, 3.62, and 3.69, respectively, and corresponds to their respective optical band gaps. The optical band gaps for X=Na, K, Rb, and Cs, agreed well with the first direct electronic transition from top of valence band (O-p state, specifically NBO-p state) to bottom of conduction band (X-s state) along $\Gamma$-symmetry. However, for X=Li the optical band gap is due to the direct electron transition from the third band of the valence band (O-p state) to the bottom of the conduction band (Li-s state) along $\Gamma$-symmetry. Along the x-axis, the respective first prominent peaks of k$^x$ for X=Li, Na, K, Rb, and Cs are located within the region 4.50 to 6.00 eV. This corresponds to the first inter-band transition between the valence band and conduction band along A-symmetry point for LPO, and NPO, and for KPO, RPO, and CPO, the transition is along Z-symmetry.
	\par The main motive for exploring the optical properties is to determine the reflectivity and the optical absorption [see figures \ref{optical} (e) and (f)]. From the reflectivity spectra as a function of the photon energy, similar to other optical properties parameter, a highly anisotropic trending of the reflectivity curves are found for each compound. The static reflectivity (R(0)) values being lower than 0.1 along x, y, and z-axes reveals the semi-conducting behavior of the considered systems. Between 0 and 4.0 eV photon energy, each compound shows increasing reflectivity with energy and then beyond 4.0 eV, drastically fluctuating reflectivity profiles are observed. Since, the maximum reflectivity even at higher photon energy region (i.e., within the ultra-violet region) is less than 25\%, it indicates that the examined materials should be good high energy electromagnetic radiation absorber than a reflector. From the curve of absorption coefficient as a function of photon energy which is consistent with other optical properties such as $\epsilon_2$ and k, one can observe a rapid increase of absorption from $\sim$3.0 eV photon energy. For each compound, along the x, y, and z-axes the first promising optical absorption are in the region of 4.0 to 5.0 eV. Since, the	most active optical absorption for all the studied materials falls within the UV-region with $\alpha$ $>$ 1$\times$10$^5$ cm$^{-1}$, these compounds might be a potential candidates for an optical materials that can serve as UV-radiation shielding materials.          
	
	\subsection{Elastic Properties}
	For practical applicability particularly in the field of piezoelectric and TE applications, the crystalline materials’ strength which is interpreted from the elastic properties plays a key role. In this regard, we report the elastic constant (C$_{ij}$) and other mechanical properties of X$_2$PbO$_3$. Since the investigated compound existed in monoclinic phase (for X=Li, and Na), and orthorhombic phase (for X=K, Rb, and Cs), we have ten elastic constants [see table \ref{MechT-1}]. The necessary and sufficient conditions for the studied systems to become mechanically stable called Born criteria are given below:\cite{Born1940e,Mouhat2014f}
	\par For monoclinic phase, the criteria are:
	\begin{equation}
	\begin{split}
	C_{11}, C_{22}, C_{33}, C_{44}, C_{55}, C_{66}>0\\
	C_{44}C_{66}-2C_{46}>0\\
	C_{11}+C_{22}+C_{33}+2(C_{12}+C_{13}+C_{23})>0\\
	C_{22}+C_{33}-2C_{23}>0
	\end{split}
	\label{EPQ-1} 
	\end{equation}
	\begin{table*}[t!]
		\small
		\caption{Calculated elastic constants $C_{ij}$ (in GPa) of X$_2$PbO$_3$ (X=Li, Na, K, Rb, Cs) using PBE0 functional.}	
		\label{MechT-1}\renewcommand{\arraystretch}{1.55}
		\begin{tabular*}{0.90\textwidth}{@{\extracolsep{\fill}}|l|llllllllll|}
			\hline
			X &  C$_{11}$ & C$_{22}$ & C$_{33}$ & C$_{44}$ & C$_{55}$ & C$_{66}$ & C$_{12}$ & C$_{13}$ & C$_{23}$ & C$_{46}$\\
			\hline
			Li &  155.99 & 202.51 & 230.62 & 67.19 & 47.69 & 54.42 & 50.38 & 34.38 & 67.17 & -11.63\\
			Na & 136.52 & 170.85 & 183.30 & 70.52 & 49.14 & 59.95 & 46.12 & 40.96 & 59.79 & 8.06 \\
			K  & 83.28 & 57.12 & 131.37 & 25.30 & 9.52 & 16.44 & 32.04 & 13.19 & 31.84 & 0.00 \\
			Rb & 74.45 & 63.81 & 117.98 & 26.76 & 7.17 & 13.32 & 31.59 & 11.35 & 31.19 & 0.00 \\
			Cs & 74.09 & 62.86 & 102.86 & 26.99 & 2.52 & 13.78 & 30.15 & 16.55 & 26.87 & 0.00 \\
			\hline
		\end{tabular*}
	\end{table*}
	
	\begin{table*}[t!]
		\small
		\caption{Elastic moduli (Bulk modulus (B), Young's modulus (Y), and Shear modulus (G) all in GPa unit), and Poisson's ratio ($\nu$) (unitless) using PBE0 functional. Here, the subscripts V, R and, H represent Voigt, Reuss and Hill assumptions, respectively.}
		\label{MechT-2}\renewcommand{\arraystretch}{1.55}
		
		\begin{tabular*}{1.00\textwidth}{@{\extracolsep{\fill}}|l|llllllllllll|}
			\hline
			X	& B$_{V}$ & B$_{R}$ & B$_{H}$ & Y$_{V}$ & Y$_{R}$ & Y$_{H}$ & G$_{V}$ & G$_{R}$ & G$_{H}$ & $\nu_{V}$ & $\nu_{R}$ & $\nu_{H}$ \\
			\hline
			Li & 99.22 & 87.11 & 93.17 & 156 & 132.71 & 144.36 & 63.01 & 53.25 &  58.13 & 0.24 & 0.25 & 0.24 \\
			Na & 87.16 & 80.48 & 83.82 & 144.10 & 134.08 & 139.09 & 58.84 & 54.85 & 56.84 & 0.22 & 0.22 & 0.22 \\
			K  &  47.32 & 44.56 & 45.94 & 59.89 & 45.85 & 52.95 & 23.23 & 17.26 & 20.24 & 0.29 & 0.33 & 0.31 \\
			Rb & 44.95 & 43.63 & 44.29 & 55.83 & 40.16 & 48.14 & 21.59 & 14.91 & 18.25 & 0.29 & 0.35 & 0.32 \\
			Cs & 42.98 & 42.26 & 42.62 & 51.37 & 23.86 & 38.14 & 19.74 & 8.49 & 14.11 & 0.30 & 0.4 & 0.35 \\
			\hline
		\end{tabular*}
	\end{table*}
	
	\begin{table*}[bth!]
		\small
		\caption{Pugh's ratio (k) (unitless) in Hill's approximation, velocity of sound (v) (in km s$^{-1}$), density ($\rho$) (in g cm$^{-3}$), Kleinman coefficient ($\zeta$) (unitless), anisotropic factor (A$_{an}$) (unitless), machinable factor ($\mu_m$) (unitless), melting temperature (T$_m$) (in K), Debye temperature ($\Theta$) (in K), and Frantsevich ratio (G/B). Here, the subscripts $t$, $l$ and $av$ represent transverse, longitudinal and average velocities, respectively.}
		
		\label{MechT-3}\renewcommand{\arraystretch}{1.55}
		
		\begin{tabular*}{0.9\textwidth}{@{\extracolsep{\fill}}|l|lllllllllll|}
			\hline
			X	& k$_{H}$ & v$_{t}$ & v$_{l}$ & v$_{av}$ & $\rho$ & $\zeta$ & A$_{an}$ & $\mu_m$ & T$_m$$\pm$300 & $\Theta$ & G/B\\ 
			\hline
			Li	& 1.60 & 2.89 & 4.86 & 3.20 & 6.94 & 0.47 & 1.96 & 1.39 & 1475.06 & 432.07 & 0.62\\
			Na	& 1.47 & 2.95 & 4.94 & 3.26 & 6.53 & 0.48 & 2.29 & 1.19 & 1359.97 & 415.51 & 0.65\\
			K	& 2.27 & 2.04 & 3.87 & 2.28 & 4.87 & 0.52 & 1.77 & 1.82 & 1045.27 & 254.71 & 0.43\\
			Rb	& 2.43 & 1.78 & 3.46 & 1.99 & 5.74 & 0.56 & 1.75 & 1.65 & 993.07  & 216.38 & 0.41\\
			Cs	& 3.02 & 1.50 & 3.12 & 1.69 & 6.29 & 0.54 & 2.06 & 1.58 & 990.95  & 177.17 & 0.33\\
			\hline
		\end{tabular*}
	\end{table*}
	
	\par For orthorhombic phase, the criteria are:
	\begin{equation}
	\begin{split}
	C_{11}, C_{22}, C_{33}, C_{44}, C_{55}, C_{66}>0\\
	C_{11}C_{22}>C_{12}^2\\
	[C_{11}C_{22}C_{33}+2C_{12}C_{13}C_{23}-C_{11}C_{23}^2\\
	-C_{22}C_{13}^2-C_{33}C_{12}^2]>0
	\end{split}
	\label{EPQ-2} 
	\end{equation}
	\par Since, the calculated elastic constants C$_{ij}$ of X$_2$PbO$_3$ compounds meet the above mentioned stability criteria, they are mechanically stable. In table \ref{MechT-1}, the C$_{11}$, C$_{22}$, and C$_{33}$ determine the system's stiffness in relation to fundamental stresses while C$_{44}$, C$_{55}$, and C$_{66}$ give their struggle against shear deformation. From the constants C$_{11}$, C$_{22}$, and C$_{33}$ $>>$ C$_{44}$, C$_{55}$, and C$_{66}$ it can be understood that the compounds X$_2$PbO$_3$ tend towards greater resistant to axial compression than shear deformation, which is re-confirmed from the fact that bulk modulus (B) are greater than shear modulus (G). Also, the high dissimilarity values of C$_{ij}$ reveals that each compound possesses anisotropic single-crystal elastic behavior. Additionally, our results of elastic constants also suggested that among the investigated X$_2$PbO$_3$ materials, the C2/c-X$_2$PbO$_3$ show greater anisotropic elastic nature compared to the C$mc$2$_1$-X$_2$PbO$_3$. Herein, the elastic moduli: the bulk modulus (B), Young's modulus (Y) and shear modulus (G) given in table \ref{MechT-2} are determined in terms of Voigt, Reuss, and Hill assumptions, which measure the uniform strain, uniform stress, and their average, respectively.\cite{Voigt1966b,Reuss1929d,Hill1952d} The larger atomic size of X as it goes from Li $\rightarrow$ Na for C2/c structure and from K $\rightarrow$ Rb $\rightarrow$ Cs for C$mc$2$_1$ structure reduces the incompressibility and resistance to volume changes due to the increasing lattice parameters which subsequently increases the inter-atomic distances, consequently, causes B to reduce. In a similar vein, the resulting Y and G likewise decrease correspondingly. This implies that the studied compounds are soft, flexible, or easier to stretch as X moves down the group. Analyzing the brittleness or ductility based on calculations of Poisson's ($\nu$) (estimated using Voigt, Reuss, and Hill assumptions) and Pugh's ratio (k) (estimated by Hill assumption only) evaluated using equation \ref{EPQ-3} and tabulated in table \ref{MechT-2} and \ref{MechT-3}, indicates that the C2/c-X$_2$PbO$_3$ are brittle in nature, while the C$mc$2$_1$-X$_2$PbO$_3$ are ductile. The ductility behavior is in the order of	CPO $>$ RPO $>$ KPO $>$ (k=1.75 or $\nu$=0.28) $>$ LPO $>$ NPO, where k=1.75 and $\nu$=0.28 are the critical values for Pugh's ratio and Poisson's ratio such that the materials become ductile.
	\begin{equation}
	\begin{split}
	\nu=\frac{Y}{2G}-1\\
	k=\frac{B}{G}
	\end{split}
	\label{EPQ-3} 
	\end{equation}
	\par The information about internal deformation stability and anisotropic factors can be gained through the Kleinman coefficient ($\zeta$) and elastic anisotropic factor (A$_{an}$) calculated using equation \ref{EPQ-4} and presented in table \ref{MechT-3}.\cite{Kleinman1962d,Chung1968b} The $\zeta$ which is in the range 0$\le$$\zeta$$\le$1 represents stretching and bending of bonds, closer value of $\zeta$ to 1 indicates a negligible contribution of bond stretching. Clearly, for C2/c-X$_2$PbO$_3$, the mechanical strengths are mostly contributed by bond stretching while, for C$mc$2$_1$-X$_2$PbO$_3$ it is mainly due to the contribution from  bending of bonds. Since, the obtained A$_{an}$ is larger than 1 for all the considered systems, they are highly anisotropic in nature (A$_{an}$=1 represents isotropic). 
	\begin{equation}
	\begin{split}
	\zeta=\frac{C_{11}+8C_{12}}{7C_{11}+2C_{12}}\\
	A_{an}=\frac{4C_{11}}{C_{11}+C_{33}-2C_{13}}	
	\end{split}
	\label{EPQ-4}
	\end{equation}
	\par In this novel work, the main concern for calculating elastic properties is to check the material's average sound velocity (v$_{av}$) determined from the transverse and longitudinal velocities (v$_t$ and v$_l$), and the Debye temperature ($\Theta$) calculated using equations \ref{EPQ-5} and \ref{EPQ-6}:\cite{Cahill1989e,Tani2010e,Chen2011h} 
	\begin{equation}
	v_{av}=\Big[\frac{1}{3}\Big(\frac{1}{{v_l}^3}+\frac{2}{{v_t}^3}\Big)\Big]^{\frac{-1}{3}}
	\label{EPQ-5}
	\end{equation}
	\par Where, v$_l$=$\sqrt{\frac{3B+4G}{3\rho}}$ and v$_t$=$\sqrt{\frac{G}{\rho}}$. Here, $\rho$ is density.
	\begin{equation}
	\Theta=\frac{h}{k}\Bigg(\frac{3nN_{A}\rho}{4\pi M}\Bigg)^{\frac{1}{3}}v_{av}
	\label{EPQ-6}
	\end{equation} 
	\par Where, h is the Planck's constant, k Boltzmann constant, n number of atoms per formula unit, N$_A$ Avogadro's number and M molecular mass.
	\par From the results presented in table \ref{MechT-3}, we observe a decreasing v$_{av}$ and $\Theta$ when X moves from Li $\rightarrow$ Na $\rightarrow$ K $\rightarrow$ Rb $\rightarrow$ Cs which is mainly due to the increasing atomic masses. For the C$mc$2$_1$-X$_2$PbO$_3$ structures, the reported v$_{av}$ are lower than those analogous compounds such as Na$_2$SiO$_3$, and Na$_2$GeO$_3$ with v$_{av}$=4.02 Km s$^{-1}$ and 3.19 Km s$^{-1}$, respectively.\cite{Zosiamliana2022h} As sound velocities are inversely proportional to the density [see equation \ref{EPQ-5}], the larger values of $\rho$ for the investigated compounds have resulted in the slower v$_{av}$ when compared with the Na$_2$SiO$_3$, and Na$_2$GeO$_3$ glass-like materials.\cite{Zosiamliana2022h,Zosiamliana2022g,Renthlei2023d} The Debye temperature being higher for C2/c-X$_2$PbO$_3$ than the C$mc$2$_1$-X$_2$PbO$_3$, suggested more number of active phonon modes in the C2/c-X$_2$PbO$_3$.
	\par For the application of any compounds in practice, particularly in the area of piezoelectric and TE applications; machinable factor ($\mu_m$), Frantsevich ratio (G/B), and melting temperature (T$_m$) played a key role.\cite{Lalrinkima2021d} The calculated $\mu_m$ by employing equation \ref{EPQ-7} suggested that the investigated X$_2$PbO$_3$ compounds exhibit an acceptable level of machinability with lower feed forces and intermediate lubricating properties which makes them a potential candidate for piezoelectric materials. The high T$_m$ and low Frantsevich ratio, also reveal that these compounds could be a future TE materials which can be utilized at high temperatures.
	\begin{equation}
	\begin{split}
	\mu_m=\frac{B}{C_{44}}\\
	T_m=(553+5.91C_{11})\pm300
	\end{split}
	\label{EPQ-7}
	\end{equation} 
	\subsection{Thermodynamic Properties}
	\begin{figure*}[t!]
		\includegraphics[height=10.0cm]{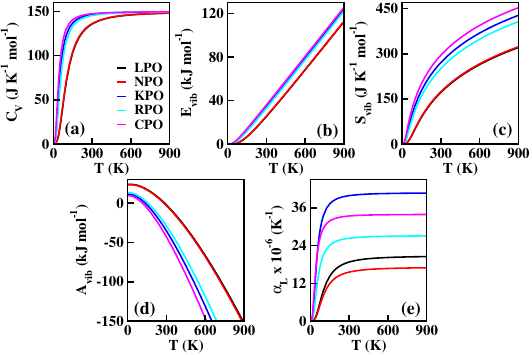}
		\caption{Calculated thermodynamics properties using GGA formalism: (a) constant-volume heat capacity C$_V$ (J K$^{-1}$ mol$^{-1}$), (b) change in vibrational internal energy E$_{vib}$ (kJ mol$^{-1}$), (c) vibrational entropy S$_{vib}$ (J K$^{-1}$ mol$^{-1}$), (d) vibrational Helmholtz free energy A$_{vib}$ (kJ mol$^{-1}$), and (e) linear thermal expansion coefficient $\alpha$ (K$^{-1}$) as a function of temperature.}
		\label{gga-thermo}
	\end{figure*}
	Knowledge of the state of any systems in terms of their energy can be gained through thermodynamic properties calculations. In this regards, we have reported the properties for X$_2$PbO$_3$ such as the constant volume specific heat (C$_V$), change in vibrational internal energy (E$_{vib}$), vibrational entropy (S$_{vib}$), vibrational Helmholtz free energy (A$_{vib}$), and the linear thermal expansion coefficient ($\alpha_L$) as a function of temperature based on the quasi-harmonic Debye model.\cite{Baroni2010a,Blanco2004d,Palumbo2017d,Palumbo2017e} From this model, 
	\begin{equation}
	\begin{split}
	C_V=3nk\Big[4D\Big(\frac{\Theta}{T}\Big)-\frac{3\Theta/T}{e^{\Theta/T}-1}\Big]\\
	S_{vib}=nk\Big[4D\Big(\frac{\Theta}{T}\Big)-3\ln\big(1-e^{-\Theta/T}\big)\Big]\\
	A_{vib}=nkT\Big[\frac{9}{8}\frac{\Theta}{T}+3\ln\big(1-e^{-\Theta/T}\big)-D\big(\Theta/T\big)\Big]\\
	\alpha_L=\frac{\gamma C_v}{B_T V}	
	\end{split}
	\label{TPE-1}
	\end{equation}
	\par Here, D is the Debye integral, $\gamma$ Gruneisen parameter,\cite{Holt1970e} and B$_T$ isothermal bulk modulus define as,
	\begin{equation}
	\begin{split}
	D(y)=\frac{3}{y^3}\int_{0}^{y}\frac{x^3}{e^x-1}dx\\	
	\gamma=\frac{3(1+\nu)}{2(2-3\nu)}\\
	B_T=-V\big(\frac{\delta P}{\delta V}\big)_T
	\end{split}
	\label{TPE-2}
	\end{equation}
	\par Where, $\nu$ is the Poisson's ratio.
	\par The intrinsic disorderliness which provides the degree of randomness within the studied systems can be determined from the study of vibrational entropy (S$_{vib}$). From figure \ref{gga-thermo} (c), it is comprehensible that the C2/c-X$_2$PbO$_3$ has lower S$_{vib}$ than those C$mc$2$_1$-X$_2$PbO$_3$. This reveals that in C$mc$2$_1$-X$_2$PbO$_3$, the amount of thermodynamic potentials (thermal energy) available for doing useful works are higher than the C2/c-X$_2$PbO$_3$. The deviation in S$_{vib}$ between the monoclinic and orthorhombic structures of X$_2$PbO$_3$ is mainly due to the variation in structural and atomic arrangement, and the change in X-atom component. Also, the increasing S$_{vib}$ for each compounds with T suggest more vibrational states become available or accessible when temperature escalates. In C2/c-X$_2$PbO$_3$, since the arrangement of atoms especially the bonding between alkali atoms with BO and NBO are identical, a nearly equivalent S$_{vib}$ curves are obtained. Whereas, for C$mc$2$_1$-X$_2$PbO$_3$, X bonding with BO and NBO are in different arrangements which leads to different internal energy contributions, thus showing a distinct thermodynamic properties. Additionally, the values of A$_{vib}$ [see figure \ref{gga-thermo} (d)] for C$mc$2$_1$-X$_2$PbO$_3$ being lower than C2/c-X$_2$PbO$_3$ re-confirms the higher availability of thermal energy for doing useful work in C$mc$2$_1$-X$_2$PbO$_3$ structures. In figure \ref{gga-thermo} (b), a linearly increasing E$_{vib}$ with T for both C2/c and C$mc$2$_1$-X$_2$PbO$_3$ are observed. This is due to the gradual rising of atom's kinetic energy with the continuous thriving heat, which leads to more atomic vibrations. Thus, subsequently results in continuous escalation of vibrational internal energy.
	\par Comprehension of lattice vibrational characteristics comes from the study of specific heat (C$_V$) which is one of the focal point for thermodynamic properties investigation. The C$_V$ curves as a function of temperature for X$_2$PbO$_3$ materials are presented in figure \ref{gga-thermo} (a). Evidently, the curves of C$_V$ showed similar trending to the nature of other thermodynamic properties parameters plot mentioned above. The higher value of C$_V$ for C$mc$2$_1$-X$_2$PbO$_3$ reveals more heat can be stored with a small gradational in the temperature compared to the C2/c-X$_2$PbO$_3$. At low temperatures (T $<<$ $\Theta$), the C$_V$ curves for each compound vary proportionally to T$^3$, i.e., it follows the Debye's T$^3$ law [see equation \ref{CV1}],    
	\begin{equation}
	C_V=\frac{12}{5}\pi^4nR\Big(\frac{T}{\Theta}\Big)^3
	\label{CV1}
	\end{equation}
	However, at high temperatures (T $>>$ $\Theta$), the C$_V$ tends to Dulong–Petit limit [see equation \ref{CV2}], 
	\begin{equation}
	C_V\simeq3nR
	\label{CV2}
	\end{equation}  
	\par Where R=8.314 J K$^{-1}$ mol $^{-1}$ is Universal gas constant. 
	\par An analysis of the linear thermal expansion coefficient ($\alpha_L$) which is closely related to the Young's modulus estimated in Voigt assumption (uniform strain assumption) yield the information about the relationship between the strain that any material endures as a temperature varies. Evidently, from $\alpha_L$ plot presented in figure \ref{gga-thermo} (e), one can find that the expansion experiences by the C2/c-X$_2$PbO$_3$ is nearly twofold to threefold lower than the C$mc$2$_1$-X$_2$PbO$_3$ compounds. The fact that Young's modulus is lower in C$mc$2$_1$-X$_2$PbO$_3$ than C2/c-X$_2$PbO$_3$ is a key factor for obtaining higher $\alpha_L$ for C$mc$2$_1$-X$_2$PbO$_3$. Comparison of thermodynamics parameters (S$_{vib}$, C$_V$ and $\alpha_L$) for the investigated compounds with some other glass-like materials such as Na$_2$SiO$_3$ and Na$_2$GeO$_3$ are listed in table \ref{TPT-1}. Clearly, the investigated compounds are thermally more stable at room temperature since they exhibit higher S$_{vib}$ and C$_V$ than those Na$_2$SiO$_3$ and Na$_2$GeO$_3$ glass-like materials. Also, the lower expansion coefficient of X$_2$PbO$_3$ than Na$_2$GeO$_3$ reveals that the continuous thriving heat will have least effect on the studied X$_2$PbO$_3$ when compared with Na$_2$GeO$_3$.  
	
	\begin{table*}[hbt!]
		\small
		\caption{\ Comparison table of S$_{vib}$ (in J K$^{-1}$ mol$^{-1}$), C$_V$ (in J K$^{-1}$ mol$^{-1}$), and $\alpha_L$ (in K$^{-1}$) at T=300 K for X$_2$PbO$_3$ (current work), Na$_2$SiO$_3$ and Na$_2$GeO$_3$.}
		
		\label{TPT-1}\renewcommand{\arraystretch}{1.55}
		\begin{tabular*}{0.60\textwidth}{@{\extracolsep{\fill}}|l|lll|}
			\hline
			\multicolumn{1}{l}{}& \multicolumn{2}{l}{This work} &\\ 
			\hline
			Compounds	 & S$_{vib}$ & C$_V$  & $\alpha_L$ \\\hline
			C2/c-Li$_2$PbO$_3$       & 167.26 & 134.59 & 18.55$\times$10$^{-6}$\\
			C2/c-Na$_2$PbO$_3$       & 170.72 & 135.77 & 15.35$\times$10$^{-6}$\\
			C$mc$2$_1$-K$_2$PbO$_3$  & 265.76 & 145.77 & 39.59$\times$10$^{-6}$\\
			C$mc$2$_1$-Rb$_2$PbO$_3$ & 248.48 & 144.59 & 26.21$\times$10$^{-6}$\\
			C$mc$2$_1$-Cs$_2$PbO$_3$ & 291.68 & 146.35 & 33.18$\times$10$^{-6}$\\
		\hline
			\multicolumn{1}{l}{} & \multicolumn{2}{l}{Other's work} &\\
			\hline
			C$mc$2$_1$-Na$_2$SiO$_3$\cite{Belmonte2016c}& 112.50 & 109.50 & -- \\
			C$mc$2$_1$-Na$_2$GeO$_3$\cite{Renthlei2023d}& 163.50 & 133.90 & 2.25$\times$10$^{-5}$ \\
			\hline
		\end{tabular*}
	\end{table*}
	
	\subsection{Piezoelectric and Electromechanical Coupling Properties}

	\begin{table*}[hbt!]
		\small
		\caption{Total direct (e$_{i\nu}$ in C m$^{-2}$ unit) and converse (d$_{i\nu}$ in pm V$^{-1}$ unit) piezoelectric properties of X$_2$PbO$_3$ (X=K, Rb, Cs) using PBE0 functional computed based on a Berry-phase numerical approach.}
		\label{piezo-1}\renewcommand{\arraystretch}{1.75}
		\begin{tabular*}{0.8\textwidth}{|l|lllll|lllll|}
			\multicolumn{1}{l}{} & &&\multicolumn{2}{l}{Direct} & &&\multicolumn{1}{l}{Converse} \\
			\hline
			X & {e$_{31}$} & {e$_{32}$} & {e$_{33}$} & {e$_{24}$} & {e$_{15}$} & {d$_{31}$} & {d$_{32}$} & {d$_{33}$} & {d$_{24}$} & {d$_{15}$} \\
			\hline
			K  & 0.32 & -0.51 & 0.30 & -0.29 & 0.29 & 9.70 & -17.41 & 5.49 & -11.53 & 30.90 \\
			Rb & -0.38 & 0.45 & -0.28 & 0.20 & -0.18 & -10.63 & 14.86 & -5.24 & 7.55 & -24.71 \\
			Cs & -0.01 & -0.03 & 0.16 & 0.02 & 0.00 & -0.11 & -1.13 & 1.83 & 0.59 & 0.08 \\
			\hline
		\end{tabular*}
	\end{table*}

\begin{table*}[hbt!]

		\small
		\caption{The electronic term (e$^e_{i\nu}$), nuclear term (e$^n_{i\nu}$), total direct 'proper' piezoelectric constant (e$_{i\nu}$=e$^e_{i\nu}$+e$^n_{i\nu}$) (all in C m$^{-2}$ unit), and electromechanical (EM) coupling factor (k$_{i\nu}$) for X$_2$PbO$_3$ (X=K, Rb, Cs) using PBE0 functional computed using CPHF/KS analytical approach.}
		\label{piezo-2}\renewcommand{\arraystretch}{1.75}
		\begin{tabular*}{1.50\textwidth}{|l|lllll|lllll|lllll|lllll|}
			\multicolumn{1}{l}{}&&\multicolumn{5}{l}{Electronic term}  &\multicolumn{4}{l}{Nuclear term} & &\multicolumn{4}{l}{Total}&&\multicolumn{4}{l}{EM coupling}\\
			\hline
			X & {e$^e_{31}$} & {e$^e_{32}$} & {e$^e_{33}$} & {e$^e_{24}$} & {e$^e_{15}$} & {e$^n_{31}$} & {e$^n_{32}$} & {e$^n_{33}$} & {e$^n_{24}$} & {e$^n_{15}$} & {e$_{31}$} & {e$_{32}$} & {e$_{33}$} & {e$_{24}$} & {e$_{15}$} & {k$_{31}$} & {k$_{32}$} & {k$_{33}$} & {k$_{24}$} & {k$_{15}$}\\
			\hline
			K  & -0.02 & 0.10 & -0.08 & 0.14 & -0.05 & 0.36 & -0.62 & 0.38 & -0.43 & 0.34 & 0.34 & -0.52 & 0.30 & -0.29 & 0.30 & 0.41 & 0.63 & 0.36 & 0.36 & 0.39 \\
			Rb & 0.02 & -0.09 & 0.06 & -0.11 & 0.05 & -0.39 & 0.53 & -0.34 & 0.31 & -0.23 & -0.37 & 0.44 & -0.28 & 0.20 & -0.19 & 0.44 & 0.52 & 0.33 & 0.24 & 0.23 \\
			Cs & -0.01 & 0.03 & -0.02 & 0.03 & -0.02 & 0.28 & -0.08 & 0.61 & -0.03 & 0.00 & 0.27 & -0.05 & 0.60 & 0.01 & -0.02 & 0.31 & 0.06 & 0.69 & 0.01 & 0.02 \\
			\hline
		\end{tabular*}
	\end{table*}

	Due to the green method of energy conversion, the piezoelectricity have attracted attention among researchers. Piezoelectric materials may prove to be a viable and sustainable source of energy if an efficient device of high mechanical stress to atomic scale polarization can be discovered. As mentioned in Section \ref{Introduction} and \ref{Electronic Properties}, due to the centrosymmetric nature of C2/c-X$_2$PbO$_3$, the compounds LPO, and NPO do not possess piezoelectricity. Therefore, in this section, we will be focusing on the piezoelectric properties for the C$mc$2$_1$-X$_2$PbO$_3$ materials, calculated based on PBE0 functionals, only i.e., the most relevant functional in reproducing experimental lattice parameters for the investigated systems. In this work, piezoelectric tensors are computed using two different methodologies. Firstly, the direct and converse piezoelectric tensors are estimated through the numerical Berry phase (BP) approach which rely on the modern theory of polarization.\cite{Erba2013b} According to BP approach [see equation \ref{PZ1}],
	\begin{equation}
	e_{i\nu}=\frac{\vert e \vert}{2 \pi V}\sum_{l}a_{li}\frac{\delta \phi_l}{\delta \eta_\nu}
	\label{PZ1}	
	\end{equation} 
	\par Here, e$_{i\nu}$ is the direct piezoelectric tensor, e electron charge, V volume, a$_{li}$ is the i$^{th}$ Cartesian component of the l$^{th}$ direct lattice basis vector a$_l$, $\phi_l$ is numerical first derivatives of the BP, and $\eta_\nu$ the strain tensor. Also, i=x, y, z; $\nu$=1, 2, 3, 4, 5, 6 (1=xx, 2=yy, 3=zz, 4=yz, 5=xz, 6=xy). 
	\par The direct and converse piezoelectric tensors are again related by the formula [see equation \ref{PZ2}]:
	\begin{equation}
	\begin{split}
	e=dC\\
	d=eS
	\end{split}
	\label{PZ2}	
	\end{equation}
	\par Secondly, the piezoelectric tensors through an analytical approach based on the Coupled Perturbed Hartree-Fock/Kohn-Sham (CPHF/KS) scheme.\cite{Baima2016,Erba2016} Herein, in addition to the electronic term's analytical computation, the internal-strain tensor is used to assess the nuclear-relaxation contribution instead of atomic coordinates numerical geometry optimizations at strained configurations. The employed equation [see equation \ref{PZ3}] for determining nuclear-relaxation contribution is:
	\begin{equation}
	e^n_{i\nu}=\frac{-1}{V_0}\sum_{aj}Z^*_{i,aj}\Gamma_{aj,\nu}
	\label{PZ3}	
	\end{equation}
	\par Here, $\Gamma_{aj,\nu}$, and Z$^*$ are the displacement-response internal-strain tensor which describes first order atomic displacements as induced by a first order strain, and the tensors containing the Born dynamical effective charges, respectively [see equation \ref{PZ4}]:
	\begin{equation}
	\begin{split}
	Z^*_{i,aj}=\frac{\delta^2 E}{\delta \xi_i\delta u_{aj}}\bigg|_\eta\\
	\Gamma_{aj,\nu}=\frac{-\delta u_{aj}}{\delta \eta_\nu}\bigg|_\xi
	\end{split}
	\label{PZ4}	
	\end{equation}
	\par Where, u$_{aj}$ are Cartesian components of the displacement vector u$_a$ of atom a (j=x, y, z).
	\par The calculated piezoelectric tensors for C$mc$2$_1$-X$_2$PbO$_3$ are presented in table \ref{piezo-1} for BP numerical approach and table \ref{piezo-2} for CPHF/KS analytical approach, respectively. The direct (e$_{i\nu}$) and converse (d$_{i\nu}$) piezoelectricity presented in table \ref{piezo-1} determine the change of polarization under a finite strain and the strain induced by an applied electric field. At first glance, it can be noticed that from table \ref{piezo-1}, the maximum direct piezoelectric constant from BP approach reduces as X goes from K $\rightarrow$ Cs (magnitude only). The increasing atomic masses of X-atoms when it goes from K $\rightarrow$ Cs progressively reduces the distortion of atomic position from their equilibrium positions, resulting in the dis-amplification of polarization tensor which reduces the piezoelectric constants. For X=K, and Rb maximum responses are e$_{32}$=-0.51, and 0.45 C m$^{-2}$ with response direction along z-axis due to strain $\eta_{yy}$. Here, the negative sign (i.e., for X=K) indicates a compressive strain that leads $\eta_{yy}$ to negative value. While for X=Cs maximum response is e$_{33}$=0.16 C m$^{-2}$, which is in z-axis direction and due to $\eta_{zz}$ strain. 
	\par It is well-known that performing piezoelectric properties calculations through an analytical fashion of internal-strain tensor of energy second-derivatives with respect to atomic displacements and lattice deformations in combination with the inter-atomic force constant Hessian matrix i.e., CPHF/KS approach shows better accuracy than performing the calculation through numerical geometry optimizations to relax atomic positions at actual strained lattice configurations i.e., BP approach. Therefore, a more detail calculations including the electronic (e$^e_{i\nu}$) and nuclear (e$^n_{i\nu}$) terms, and the total direct 'proper' (e$_{i\nu}$) piezoelectric constants through CPHF/KS approach are reported in table \ref{piezo-2} for the C$mc$2$_1$-X$_2$PbO$_3$ compounds. Evidently, each compound's total direct 'proper' piezoelectric response contribution from the e$^e_{i\nu}$ are negligibly small and also reduces as X moves from K $\rightarrow$ Cs. The degree of polarization reduced resulting from a decrease in electronegativity down the periodic table group, which suggests that the contribution of e$^e_{i\nu}$ has diminished. Consequently, e$^n_{i\nu}$ accounts for the majority of each material's total direct piezoelectric responses. Interestingly, the total direct and total direct 'proper' piezoelectric constants obtained from BP and CPHF/KS schemes agreed well for X=K, and Rb with deviation $\vert$e$_{i\nu}$$\vert$ $<$ 0.02 C m$^{-2}$. However, in case of X=Cs, a comparatively high piezoelectric constants appears for e$_{31}$, and e$_{33}$ when CPHF/KS is adopted. This could be due to the more robust approach of CPHF/KS scheme during e$^n_{i\nu}$ calculations where partition of the nuclear-relaxation term of the piezoelectric tensor into phonon-mode contributions is taken into consideration. Overall, the maximum piezoelectric constants attributed by each compound are higher than the $\alpha$-quartz, a standard piezoelectric material, whose largest constant e$_{11}$ is 0.15 C m$^{-2}$ at room temperature and 0.07 C m$^{-2}$ at 5 K.\cite{Tarumi2007c} Also, compared to the analogous compounds such as Na$_2$SiO$_3$ and Na$_2$GeO$_3$, our results show better responses than Na$_2$SiO$_3$ (e$_{32}$=0.22 C m$^{-2}$), however lower than Na$_2$GeO$_3$ (e$_{33}$=0.90 C m$^{-2}$).\cite{Zosiamliana2022h,Renthlei2023d}
	\par The electromechanical (EM) coupling factor (k$_{i\nu}$) calculated using equation \ref{PZE-1} are presented in table \ref{piezo-2}.\cite{Celestine2024a} It determines the efficiency of piezoelectric material i.e., factor with which the materials convert mechanical energy into electrical energy or vice versa.
	\begin{equation}
	k_{i\nu}=\frac{|e_{i\nu}|}{\sqrt{C_{44}\epsilon_0\epsilon_1(0)}}
	\label{PZE-1}
	\end{equation}
	\par Here, e$_{i\nu}$ is piezoelectric constants, C$_{44}$ elastic constant (in Pa unit), $\epsilon_0$=8.854$\times$10$^{-12}$ Fm$^{-1}$ is absolute permittivity of free space and $\epsilon_1$ (0) is real static dielectric constant.  
	\par Firstly, it can be noticed that the maximum k$_{i\nu}$ for each compound corresponds to the direction of their respective maximum piezoelectric constant. Eventhough the computed k$_{i\nu}$ shows highest value for X=Cs with k$_{33}$=0.69, the overall coupling factors when all the directions are taken into account is highest for X=K, with X=Rb serving as an intermediary. This reveals a better efficiency of electromechanical transducer between the stated electric and elastic channels for X=K, and Rb. Since, our results of maximum k$_{i\nu}$ are higher than the quartz crystal\cite{Yu2010} with highest electromechanical coupling coefficient $\sim$0.29 in a direction which makes an angle of 73$^\circ$ with y axis in yz plane, the investigated C$mc$2$_1$-X$_2$PbO$_3$ could be a potential candidate which can serve as an efficient piezoelectric materials.  
	\subsection{Thermoelectric Properties}
	\begin{figure}[hbt!]
		\includegraphics[height=10.0cm]{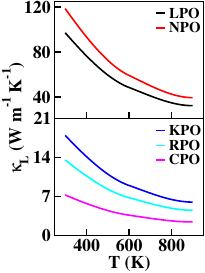}
		\caption{Calculated lattice thermal conductivity ($\kappa_L$) of X$_2$PbO$_3$ compounds using Slack equation.}
		\label{TE-1}
	\end{figure}
	\begin{figure*}[t!]
		\includegraphics[height=6.55cm]{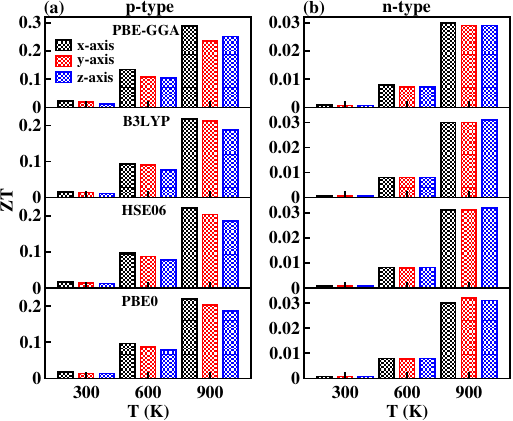}
		\includegraphics[height=6.55cm]{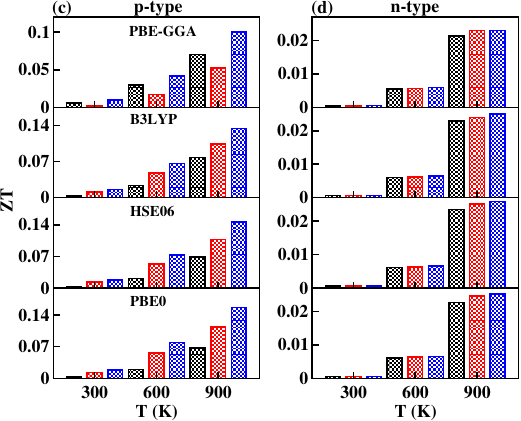}
		\includegraphics[height=6.55cm]{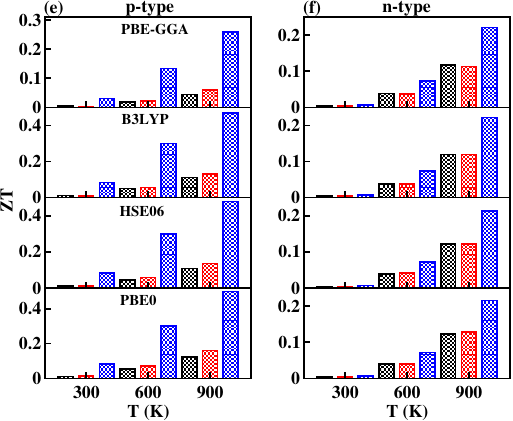}
		\includegraphics[height=6.55cm]{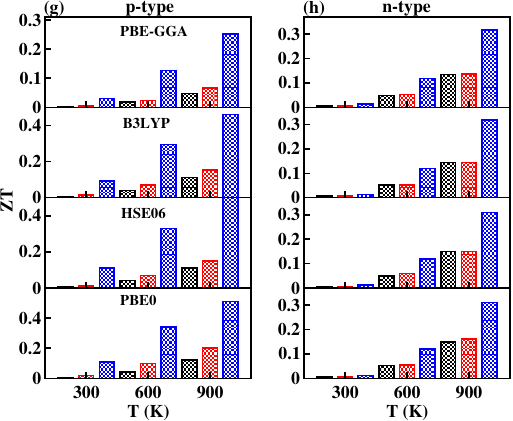}
		\includegraphics[height=6.55cm]{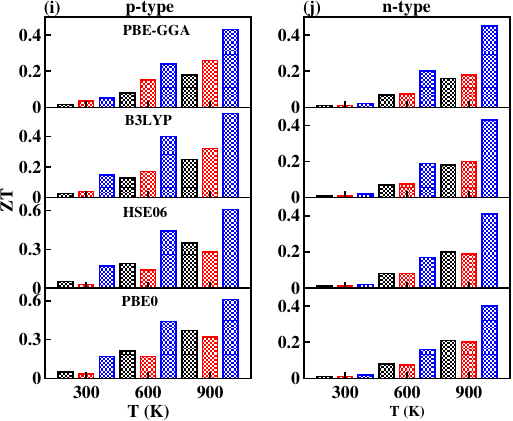}
		\caption{Calculated thermoelectric efficiency (ZT) as a function of temperature using PBE-GGA, B3LYP, HSE06, and PBE0 functionals: (a) p-type, (b) n-type for Li$_2$PbO$_3$, (c) p-type, (d) n-type for Na$_2$PbO$_3$, (e) p-type, (f) n-type for K$_2$PbO$_3$, (g) p-type, (h) n-type for Rb$_2$PbO$_3$, and (i) p-type, (j) n-type for Cs$_2$PbO$_3$.}
		\label{TE-2}
	\end{figure*}
	The prime focus of this work is to identify the TE efficiency (ZT) of the investigated compounds, which is in close relation with the electron transport properties such as Seebeck coefficient (S), electrical conductivity ($\sigma$), electron thermal conductivity ($\kappa_e$), and the lattice thermal conductivity ($\kappa_L$). For this particular calculation, we have considered not only the p-type and n-type doping, but also TE properties along the x, y, and z-axes by employing four different functionals such as PBE-GGA, B3LYP, HSE06, and PBE0. Herein, calculations for electron transport properties are performed using the Boltzmann transport equation (BTE) within the CRTA at fix $\tau$=10 fs (10$^{-14}$ s) using BoltzTraP.\cite{Madsen2006a} The employed formulae for thermoelectric parameter calculations are:\cite{Linnera2018a,Sansone2017a}   
	\begin{equation}
	\sigma(\mu,T)=e^2\int dE\bigg(\frac{-\delta f_0}{\delta E}\bigg)\Xi_{qr}(E)
	\label{Eq-TE1}
	\end{equation}
	\begin{equation}
	S(\mu,T)=\frac{e}{T}\int dE\bigg(\frac{-\delta f_0}{\delta E}\bigg)(E-\mu)\Xi_{qr}(E)
	\label{Eq-TE2}
	\end{equation}
	\begin{equation}
	\kappa_e(\mu,T)=\frac{1}{T}\int dE\bigg(\frac{-\delta f_0}{\delta E}\bigg)(E-\mu)^2\Xi_{qr}(E)
	\label{Eq-TE3}
	\end{equation}
	\par Where, $\mu$ is the chemical potential or the Fermi level, E is the energy, $f_0$ is the Fermi-Dirac distribution and $\Xi$ is the kernel of the transport distribution function (TDF). Here, $\Xi$ is further expressed as:
	\begin{equation}
	\Xi_{qr}(E)=\tau\sum_{k}\frac{1}{N_k}\frac{1}{V}\sum_{i,j}v_{i,q}(k)v_{j,r}(k)\delta(E-E_i(k))
	\label{Eq-TE4}
	\end{equation}
	\par Where, $v_{i,q}$(k) is the velocity of the $i^{th}$ band calculated along the Cartesian direction q, $\tau$ is the lifetime which is assumed to be constant according to the RTA.
	In the above equations, $\sigma$ is the electrical conductivity, S is the Seebeck coefficient, and $\kappa_e$ is the electronic thermal conductivity. 
	\par Here, the purpose for calculating the electronic transport properties is to obtain TE efficiency (figure of merit) ZT, given by
	\begin{equation}
	ZT=\frac{S^2\sigma T}{\kappa_T} 
	\label{Eq-TE5}	
	\end{equation}
	\par Where, $\kappa_T$ is the total thermal conductivity and is equal to $\kappa_e$+$\kappa_L$. To determine the lattice contribution to thermal conductivity ($\kappa_L$), we employed the well-known Slack equation given by\cite{Slack1979}
	\begin{equation}
	\kappa_L=\frac{A\bar{M}\Theta_D^3\delta}{\gamma^2Tn^{\frac{2}{3}}}
	\label{Eq-TE6}	
	\end{equation}
	\par Where, a constant A=$\frac{2.43 \times 10^{-8}}{1-0.514/\gamma+0.228/\gamma^2}$, $\bar{M}$ is average atomic mass, $\Theta_D$ is the Debye temperature, $\delta^3$ is volume per atom, $\gamma$ is the Gruneisen parameter, and n is number of atoms per unit cell.
	\par From the transport properties of electrons with respect to chemical potential ($\mu$) presented in figures S2-S14, considering the directional wise analysis, one can find that each transport properties for the investigated materials are highly anisotropic in nature with z-axis as the most preferable direction for achieving maximum TE efficiency except for LPO where it is obtained along x-axis. Also, from these plots, it can be deduced that all of the compounds under investigation exhibit p-type semi-conducting nature, with holes serving as majority carriers, regardless of the functionals employed. With hybrid functionals, each compound's E$_g$ increases which inturn rises the (E-$\mu$) term of equation \ref{Eq-TE2}. This subsequently maximizes the Seebeck coefficient (S) computed using hybrid-DFT when compared to PBE-GGA results [see figures S9, S10, and S10]. Also, on analyzing and comparing the PBE-GGA and hybrid-DFT band structure profiles of the investigated compounds [see figures \ref{band}, and S1], it is evident that the incorporated hybrid functionals have had a notable impact, particularly with regard to the energy levels where the electronic band lines are positioned. Thus, these substantial variations in band profiles due to the functionals employed has led to considerable changes in the ensuing transport properties. The resultant change in band energy due to the different functionals adopted has led to the change in density of states, being a key factor for which the term $\frac{-\delta f_0}{\delta E}$ of equations \ref{Eq-TE1}, \ref{Eq-TE2}, and \ref{Eq-TE3}, that defines the number of states mapped and offered information of the channel allowing the charges to flow, varies. This phenomenon leads to a remarkable changes of each compound's electrical conductivity ($\sigma$) and electron thermal conductivity ($\kappa_e$), when PBE-GGA and hybrid-DFT computed transport properties are compared.
	\par The primary aim for calculating transport properties is to obtain the figure of merit (ZT) which determines the material’s TE efficiency. In this regard, the TE power factor (PF=S$^2$ $\sigma$) is firstly computed that measures the TE performance of the materials. The PF with respect	to chemical potential plots are depicted in figures S12, S13, and S14. From these plots, it is revealed that the PF increased as temperature rises, suggesting the enhancement of TE efficiency of the materials with temperature. It is important to note that the first peak points of PF along the left and right sides of the zero chemical potential (Fermi energy) corresponds to the optimum chemical potentials for p-type and n-type doping at various temperatures, where the optimal ZT values are determined as a function of temperature. Obviously, the ZT [see equation \ref {Eq-TE5}] depends on the conflicting nature of $\sigma$ and $\kappa_e$,	S and also on the $\kappa_L$. Since, the transport properties from BoltzTraP does not generate $\kappa_L$, therefore an analytical Slack model given in equation \ref{Eq-TE6} is employed. The obtained $\kappa_L$ curves in figure \ref{TE-1} shows continuously decreasing plots as temperature rises due to the increasing phonon scattering. The larger atomic sizes as X goes from K$\rightarrow$Cs for C$mc$2$_1$-X$_2$PbO$_3$ results in less possibility of phonon scattering down the group. Thus, the order of $\kappa_L$ for C$mc$2$_1$-X$_2$PbO$_3$ is KPO $>$ RPO $>$ CPO. However, our observation shows a conflicting nature for C2/c-X$_2$PbO$_3$ compounds where $\kappa_L$ of NPO $>$ $\kappa_L$ of LPO. This might be due to the LPO attaining a higher density than NPO, which decreases heat transfer average distances, therefore minimizes $\kappa_L$ for LPO. Furthermore, the non-centrosymmetric distortions in the C$mc$2$_1$-X$_2$PbO$_3$ phase contribute to enhanced phonon scattering, leading to lower $\kappa_L$ values and improved thermoelectric performance relative to the centrosymmetric C2/c-X$_2$PbO$_3$ counterparts. Additionally, the calculated $\kappa_L$ values for C$mc$2$_1$-X$_2$PbO$_3$ are lower than those of the structurally related C$mc$2$_1$-Li$_2$GeO$_3$, which can be attributed to the presence of the heavy Pb atom.\cite{Zosiamliana2025} The increased atomic mass of Pb further suppresses phonon transport, reinforcing the observed reduction in $\kappa_L$. Evidently, from ZT plot as a function of temperature shown in figure \ref{TE-2}, it can be seen that each compound's ZT value improves as temperature increases. Furthermore, it can be determined that CPO has attained the best TE efficiency with ZT=0.61 at T=900 K along the z-direction for p-type doping with PBE0 functional. Interestingly, each material's computed ZT from hybrid-DFT shows better efficiency when compared to PBE-GGA result, except for LPO where PBE-GGA estimates the greatest ZT values. The findings herein are quite astonishing even-though we could achieve only a mediocre TE efficiency (the benchmark value for efficient TE materials is ZT $\ge$ 1), as this work presents the first time theoretical reported TE properties of C2/c-X$_2$PbO$_3$ and, C$mc$2$_1$-X$_2$PbO$_3$ glass-like materials with an acceptable ZT values at higher temperatures. However, rigorous research is further required so as to enhanced the electrical conductivity especially for the C$mc$2$_1$-X$_2$PbO$_3$ such that these materials could be practically applicable for future TE devices.     
	\section{Conclusion}
	We present a comprehensive investigation of the multi-functional properties of X$_2$PbO$_3$ (X=Li, Na, K, Rb, and Cs) using hybrid-DFT within the frameworks of B3LYP, HSE06, and PBE0 functionals. Our study reveals that X$_2$PbO$_3$ compounds crystallize in two distinct structural phases, namely the monoclinic (C2/c) and orthorhombic (C$mc$2$_1$) symmetries, depending on the ionic radius of the X-site atoms. To ensure the practical feasibility of these materials, we conducted a series of stability assessments: (1) The room temperature stability (from MD-simulation), (2) Mechanical stability confirmed using Born's criteria, and (3) Ground-state stability or energetic stability established from formation energy calculations. The electronic band structure analysis demonstrates a significant band gap increasing within the hybrid functionals, highlighting the semiconducting nature of X$_2$PbO$_3$. From the optical property calculations, we observed a strong absorption coefficient ($\alpha$ $>$ 1$\times$10$^5$ cm$^{-1}$) within the ultraviolet (UV) region, suggesting their potential use as UV-radiation shielding materials. The piezoelectric response of the non-centrosymmetric C$mc$2$_1$-X$_2$PbO$_3$ phase was examined using both the BP approach and the CPHF/KS method. The obtained piezoelectric constants are found to be comparable to those of standard piezoelectric materials such as $\alpha$-Quartz ($\alpha$-SiO$_2$), indicating the potential of X$_2$PbO$_3$ for electromechanical (piezoelectric) applications. Furthermore, the TE analysis suggests that these compounds exhibit promising TE performance at high temperatures, making them potential candidates for waste heat recovery and energy conversion applications. Overall, this study highlights the versatility and technological potential of X$_2$PbO$_3$, paving the way for their exploration in piezoelectric, TE, and optical applications. Experimental validation is encouraged to further substantiate the predicted piezoelectric and TE behavior. Additionally, this work provides valuable insights for computational researchers in selecting optimal hybrid functionals for accurately reproducing lattice parameters in DFT-based studies. 
\section{Acknowledgement}
	\textbf{DPR} acknowledges Anusandhan National Research Foundation (ANRF), Govt. of India, vide Sanction Order No.:CRG/2023/000310, \& dated:10 October, 2024.\\
	\textbf{A. Laref} acknowledges support from the "Research Center of the Female Scientific and Medical Colleges",  Deanship of Scientific Research, King Saud University.\\


\section{Conflict of Interest}
The authors declare no competing financial interest

\section{Author contributions}
\begin{enumerate}
\item R. Zosiamliana: Performed detail calculations, Formal analysis, Visualization, Validation, Literature review, Writing-original draft, writing-review \& editing.
\item Lalhriat Zuala: ormal analysis, Visualization, Validation, writing-review \& editing. 
\item Lalrinthara Pachuau: ormal analysis, Visualization, Validation, writing-review \& editing. 
\item Lalmuanpuia Vanchhawng: ormal analysis, Visualization, Validation, writing-review \& editing.
\item S. Gurung: ormal analysis, Visualization, Validation, writing-review \& editing.
\item A. Laref:Formal analysis, Visualisation, Validation, writing-review \& editing. 	
\item D. P. Rai: Project management, Supervision, Resources, software, Formal analysis, Visualisation, Validation, writing-review \& editing.
\end{enumerate}

\section*{Data Availability Statement}
\begin{center}
	\renewcommand\arraystretch{1.2}
	\begin{tabular}{| >{\raggedright\arraybackslash}p{0.35\linewidth} | >{\raggedright\arraybackslash}p{0.6\linewidth} |}
		\hline
		\textbf{AVAILABILITY OF DATA} & \textbf{STATEMENT OF DATA AVAILABILITY}\\  
		\hline
		Data available on request from the authors and supplementary
		&
		The data that support the findings of this study are available from the corresponding author upon reasonable request.
		\\\hline
	\end{tabular}
\end{center}


\bibliographystyle{aipnum4-1}
\bibliography{aipsamp}
\end{document}